\documentclass{elsart}
\usepackage{fleqn,graphicx}
\begin{document}

\begin{frontmatter}

\title{Spin-wave spectrum of copper metaborate in the
commensurate phase $10K<T<21K$}
\author[psi,ill]{M.~Boehm}
\author[kra]{S.~Martynov}
\author[psi]{B.~Roessli}
\author[kra]{G.~Petrakovskii}
\author[ill]{J.~Kulda}
\address[psi] {Laboratory for Neutron Scattering, ETH Zurich \& Paul
Scherrer
Institute, CH-5232 Villigen PSI}
\address[kra]{Institute of Physics, SB RAS, 660036 Krasnoyarsk, Russia}
\address[ill]{Institute Laue-Langevin, 38042 Grenoble, Cedex 9, France}

\begin{abstract}

We have investigated the spin-wave spectrum of copper metaborate, CuB$_2$O$_4$, by means
of inelastic neutron scattering in the commensurate magnetic phase.
We have found two branches of spin-wave excitations associated 
with the two  magnetic sublattices Cu(A) and Cu(B), respectively.  
In the temperature regime $10K \le T \le 21K$, where only the Cu(A)
magnetic moments
are ordered, the interaction between 
the two sublattices is found to be negligible. With this 
approximation we have determined the `easy plane' exchange
parameters of the Cu(A) subsystem within standard
spin-wave theory.
\end{abstract}
\end{frontmatter}
\section{Introduction}
\begin{figure}
     \begin{center}
     \includegraphics*[width=15cm]{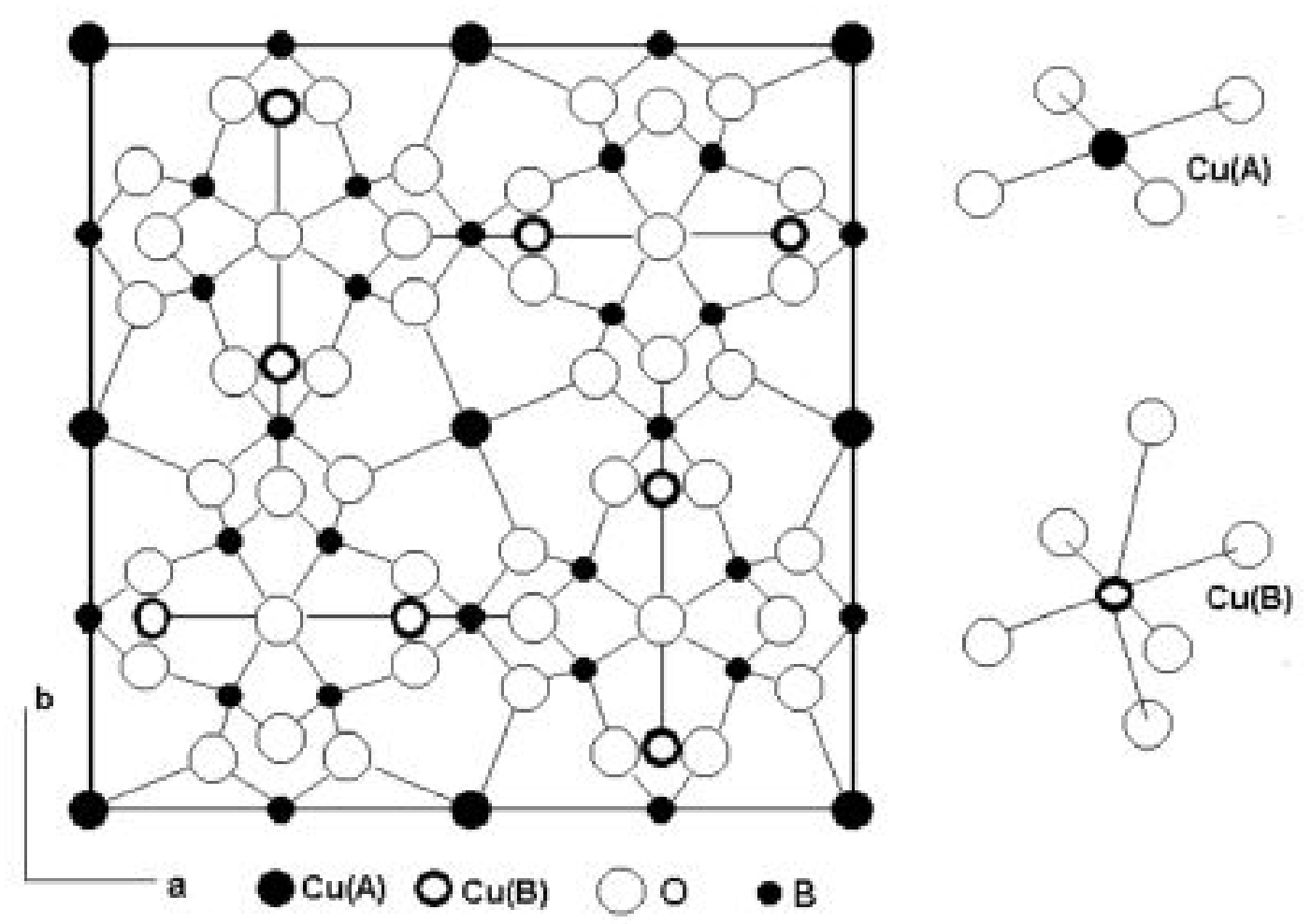}
     \end{center}
     \caption{The chemical cell of CuB$_2$O$_4$. Cu(A) and Cu(B) positions
     are represented by black and open symbols, respectively.}
     \label{fig:structure}
\end{figure}

Copper metaborate, CuB$_2$O$_4$, crystallizes with the space group $I\bar42d$ 
($D^{12}_{2d}$) with 12 formula units in the chemical cell. Although the 
chemical structure was 
already solved 30 years ago by x-ray 
measurements (\cite{martinez}, see Fig.~\ref{fig:structure}), it was only 
recently that the surprising details of the magnetic structure have been 
revealed with help of various experimental techniques
(\cite{petrakovskii1}-\cite{petrakovskii3}).
At T$_N$=21K the magnetic 
system undergoes a second order phase transition from a paramagnetic 
into a commensurate antiferromagnetic state with a magnetic propagation vector 
$\vec{k}_0=(0,0,0)$, i.~e.~the dimension of the magnetic unit 
cell is identical with the chemical one. 
Below T$^{*}$=10K the propagation vector changes continuously 
from $\vec{k}_0=(0,0,0)$ to $\vec{k}_0=(0,0,0.15)$ (r.~l.~u.)~at 

T=200mK, showing 
that the magnetic structure becomes incommensurate with respect to the 
chemical lattice along the tetragonal c-axis. Near T$^{*}$ we observed 
higher order harmonics in the diffraction pattern of neutron 
diffraction measurements~\cite{roessli,boehm}, which indicates the existence of a magnetic soliton 
lattice in this compound without the application of an external field.\par
The peculiarities of the magnetic structures are due to the existence of
two nonequivalent types of Cu $^ {2 +}$ ions, labeled 
Cu(A) and Cu(B) in the following. The Cu(A) ion is located at site $4b$ (point symmetry
$S_4,00\frac{1}{2}$) and is at the center of a square unit formed by four oxygen
ions. Cu(B) is at site $8d$  (point symmetry $C_2,x\frac{1}{4}\frac{1}{8}$, x=0.0815) 
and is surrounded by six oxygen ions located at the vertices of a distorted octahedron.
In the commensurate phase (T$ ^ *$=10K$<$T$<$T$_N$=21K), the 
Cu(A) moments are confined close to the tetragonal plane and 
the value of the moment at T=12K is nearly the same as the value of a free  
Cu $^ {2 +} $ spin of 1$\mu_B$~\cite{boehm}. In 
contrast, the value of the Cu(B) magnetic moment reaches only a value of 0.25 $\mu_B$ 
at T=12K. Hence, the magnetic state in the commensurate phase seems to be  
mainly governed by the exchange interactions of the `strong' (in the 
magnetic sense) Cu(A) subsystem.
Exchange interactions inside the second subsystem, as well 
as between the two spin systems are much weaker and become only important 
below the second transition which occurs at T*, leading to the observed changes in 
the magnetic structure.\par
In this work we analysed the spectrum of spin excitations in the 
commensurate phase in order to determine the exchange parameters 
in the strong subsystem. This entailed much simplification of the 
complex magnetic system in CuB$_2$O$_4$, but our assumptions are justified by the 
excellent agreement between experimental data and theory, as we shall  
demonstrate below.
\section{Experimental details}
\begin{figure}
     \begin{center}
     \includegraphics*[width=15cm]{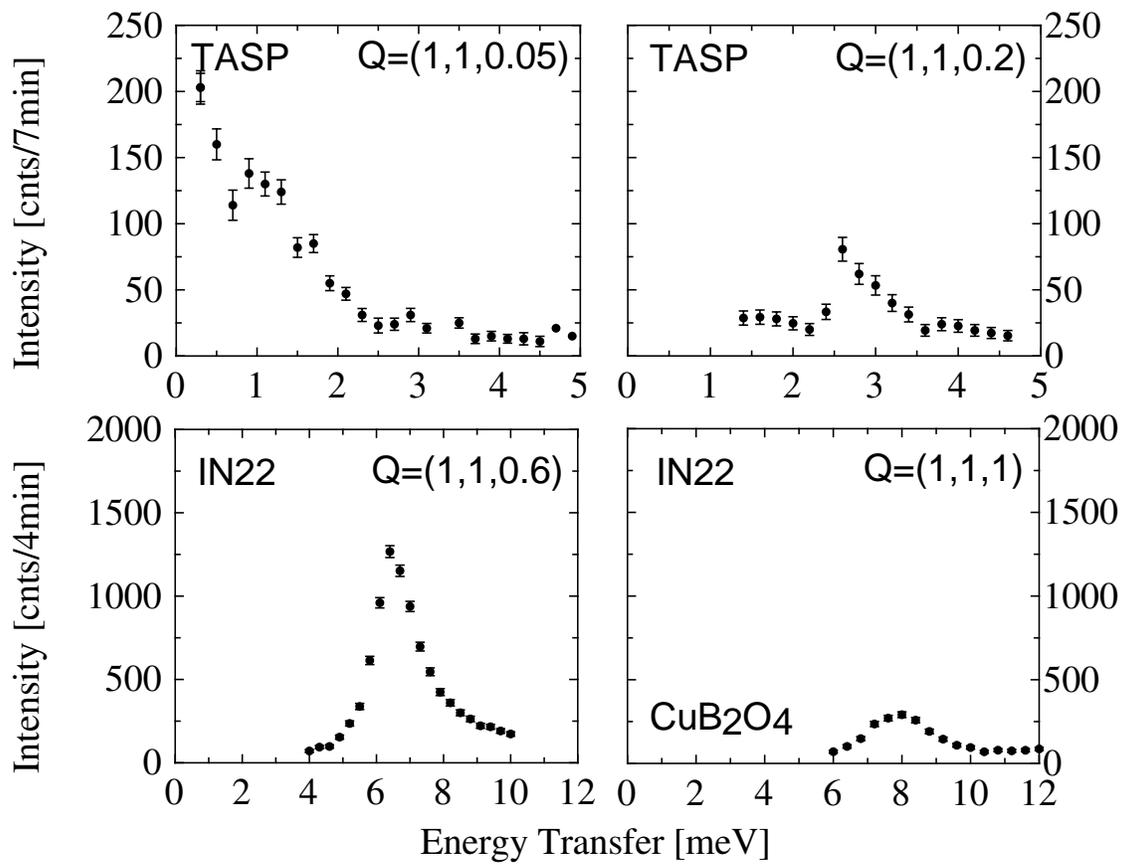}
     \end{center}
     \caption{Typical constant q-scans taken in CuB$_2$O$_4$ at T=12 K 
     at different positions in 
     reciprocal space along the (0,0,q) direction.}
     \label{fig:scans}
\end{figure}

The inelastic neutron scattering experiments were
performed at the Institut Laue Langevin (ILL), Grenoble, on the 
thermal triple-axis spectrometer IN22. 
The spectrometer was operated in the constant final
energy mode with a  neutron wave vector $k_f$=2.662 \r{A}$^{-1}$.
Contamination from higher order neutron wavelengths has been suppressed
by the use of a pyrolitic-graphite placed in the scattered beam. 
In order to increase the neutron intensity, the collimation 
was removed from the beam and the analyzer was horizontally 
focussed. In this configuration, the energy resolution at the elastic position 
was about 0.8 meV. A single crystal of CuB$_2$O$_4$ of about 0.5 cm$^3$ 
was mounted inside an $^4$He cryostat of ILL-type with the 
[q,q,0] and [0,0,q] crystallographic directions in the scattering plane. 
Typical constant q-scans taken with a counting time of $\sim$4 minutes/point 
are shown in Fig.\ref{fig:scans}. 
Additional scans were taken on the cold 
neutron triple-axis spectrometers TASP (SINQ) with a final wave-vector 
$k_f$=1.64 \r{A}$^{-1}$ and IN14 (ILL) with a final wave-vector 
$k_f$=1.55 \r{A}$^{-1}$ which yielded an improved energy resolution 
($\Delta E \sim$ 0.20 meV) and showed that 
the high energy branch of the spin-wave excitations in CuB$_2$O$_4$ 
goes to $\hbar \omega =0$ when $\vec q \to 0$. In addition, a low energy 
branch  with a 
dispersion confined below $\hbar \omega\sim$1.5 meV could be identified. 
This branch has dispersive modes along the c-axis only.     
\section{Spin-wave theory} 
\begin{figure}
     \begin{center}
     \includegraphics*[width=15cm]{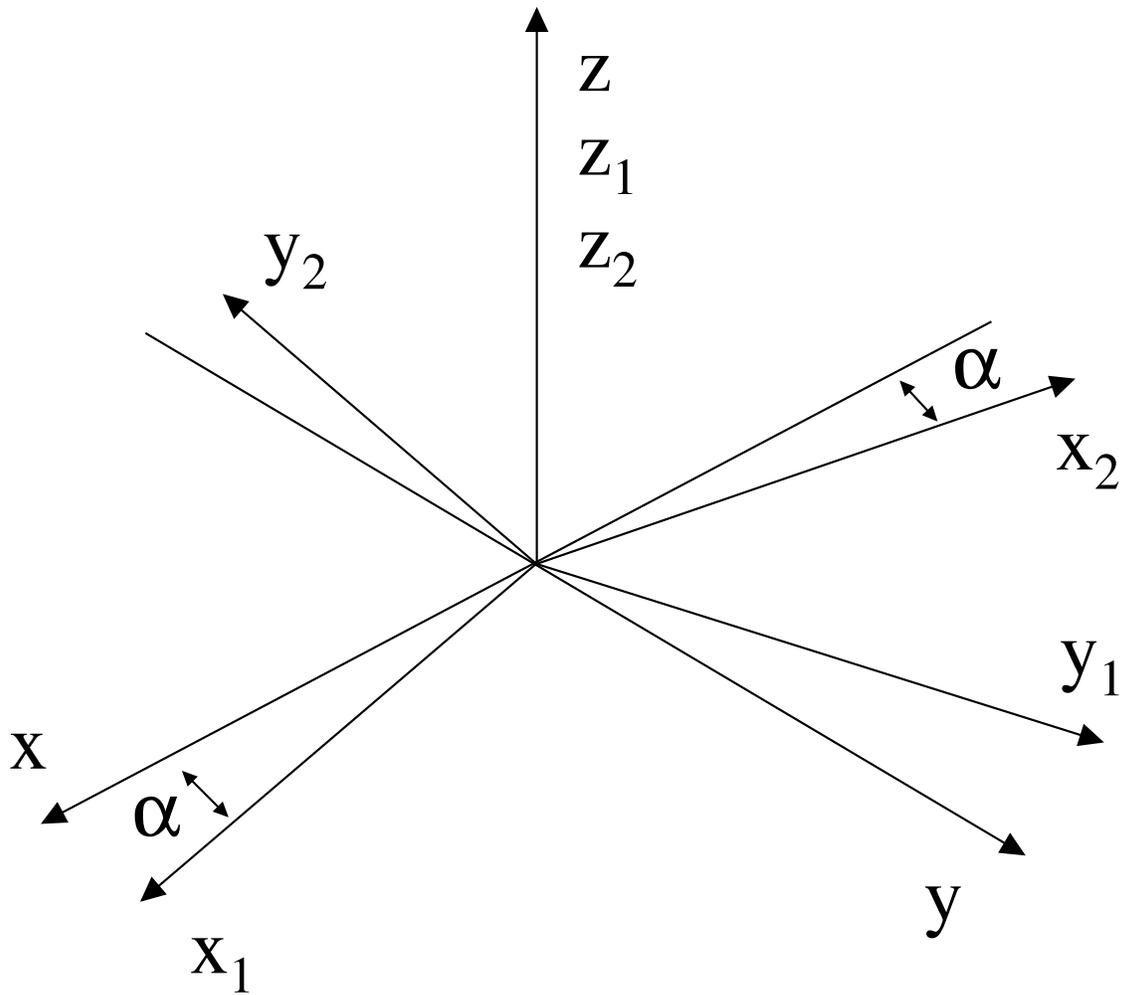}
     \end{center}
     \caption{Relation between the crystal coordinate system and the local axes
     of the two Cu(A) magnetic sublattices.}
     \label{fig:coordinates}
\end{figure}

\begin{figure}[!hbp]
     \begin{center}
     \includegraphics*[width=15cm,height=22cm]{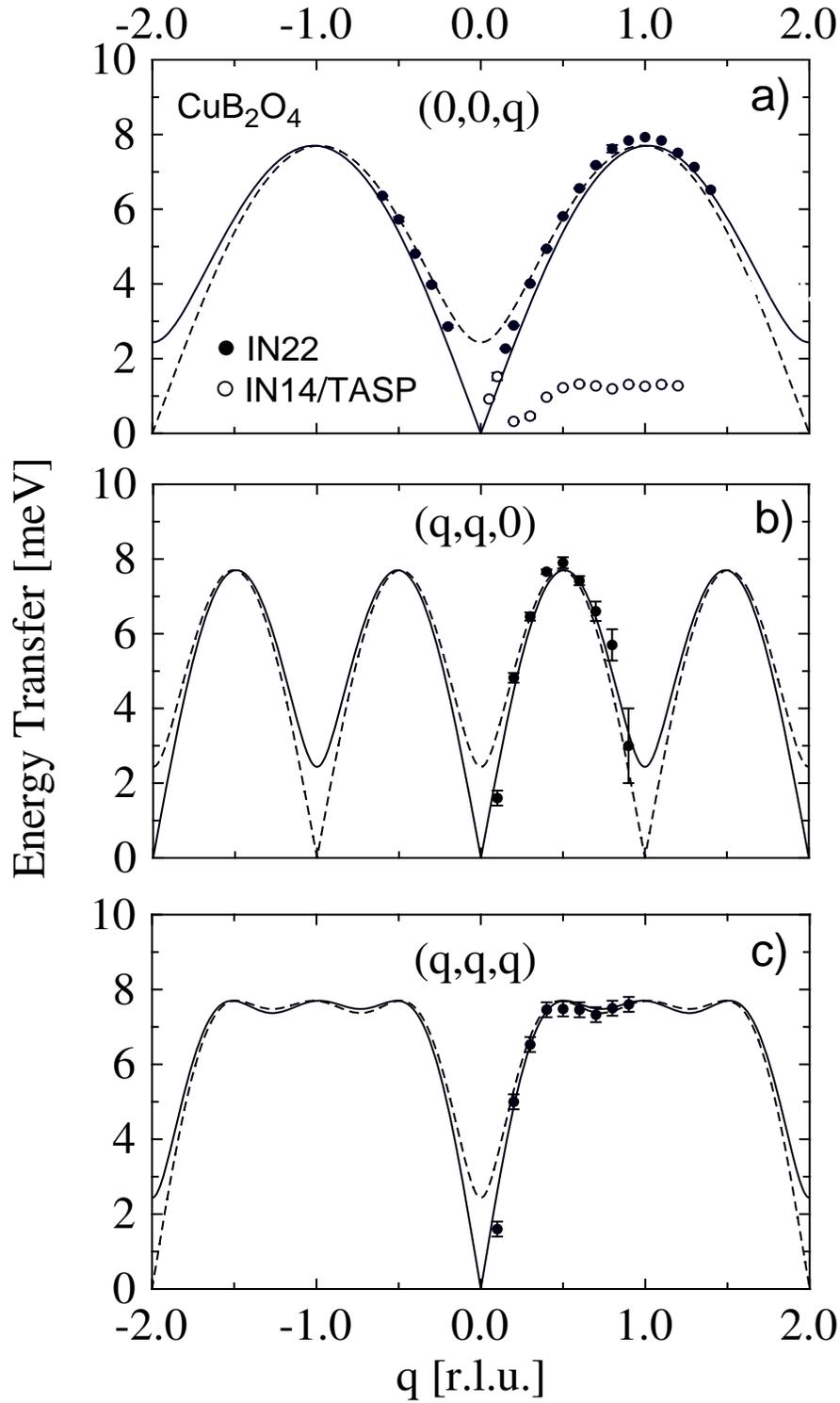}
     \caption{Spin-wave spectrum of CuB$_2$O$_4$ at T=12K. Circles - experimental data 
     points obtained from inelastic neutron scattering, solid and dash lines - 
     result of the theoretical analysis for (a)-(0,0,q), (b)-(q,q,0) and (c)-(q,q,q)
     directions, respectively.}
     \label{fig:dispersion}
     \end{center}
\end{figure}

The analysis of the upper (high energy) branches in the spin-wave spectrum
of copper metaborate can be done in a first approximation by
neglecting interactions between the different subsystems and assuming 
exchange interactions in the `strong' subsystem Cu(A) only. 
In the following, we considered the exchange interactions with four neighbour atoms 
at the positions $(0,\pm \frac{1}{2},-\frac{1}{4})$ and 
$(\pm \frac{1}{2},0,\frac{1}{4})$ relativ to the central Cu(A) ion. Interactions 
with additional neighbours can only exist in long cation chains (see 
Fig.\ref{fig:structure},\cite{martinez}) and 
do not influence the quality of the analysis of the spin-wave spectrum as shown below. 
For the same reasons we ignored other possible weak
interactions such as 
anisotropic exchanges within the basal plane and dipole-dipole interactions. 
Measurements of the magnetization~\cite{petrakovskii2}, the magnetic 
resonance~\cite{pankraz} and neutron investigations of the magnetic 
structure~\cite{roessli,boehm} have shown that
Cu(A) ions form a two-sublattice antiferromagnetic
structure with magnetic moments placed almost in the tetragonal plane ("easy
plane").  
The Hamiltonian with
S=$\frac{1}{2}$ is thus of the form
\begin{equation}
H = J_z\sum_{ij}S_i^zS_j^z+J_{xy}\sum_{ij}(S_i^xS_j^x+S_i^yS_j^y)+
h(\sum_iS_i^y+\sum_jS_j^y).
\end{equation}
The parameter $h$ plays the role of an external magnetic
field ($h=g\mu_BH $) applied in the basal plane $h=h_y$, or describes the 
Dzyaloshinskii field where $z$ is the tetragonal axis of the crystal and 
$J _ {xy} > J_z $. The Holstein-Primakoff transformation within the spin 
wave theory approach~\cite{tyablikov} for each sublattice 
($x_iy_iz_i $) and ($x_jy_jz_j $) is defined as 
\begin{eqnarray}
S_{i,j}^x=-S+b_{i,j}^+b_{i,j}\nonumber\\ 
S_{i,j}^y=\sqrt{\frac{S}{2}}(b_{i,j}^++b_{i,j})\\
S_{i,j}^z=i\sqrt{\frac{S}{2}}(b_{i,j}-b_{i,j}^+).\nonumber
\label{hp}
\end{eqnarray}
The relations between the crystal coordinate system and the local axes
are given by 
\begin{eqnarray}
x&=&x_i\cos\alpha-y_i\sin\alpha=-x_j\cos\alpha-
y_j\sin\alpha\nonumber\\
y&=&x_i\sin\alpha+y_i\cos\alpha=x_j\sin\alpha-
y_j\cos\alpha\nonumber\\
z&=&z_{i,j},\nonumber
\end{eqnarray}
as shown in Fig.~\ref{fig:coordinates}.
Using the above equations, the Hamiltonian up to the
second power in the quantization operators becomes
$$H=H_0+H_1+H_2.$$
The requirement that the linear part of the Hamiltonian, $H_1$, vanishes, leads to
a condition for the orientation of the equilibrium magnetization of 
each sublattice:
\begin{equation}
\sin\alpha=\frac{h}{2h_\bot},\nonumber
\end{equation}
where $h_\bot=nSJ_{xy}$. $n$ is the number of nearest neighbours, $H_0
=-nS^2J_{xy}N$ the energy of the ground state and N the number of spins in
one sublattice.
For the evaluation of the spin-wave spectra, one has to consider the part 
in the Hamiltonian which is quadratic in the quantization operators, i.e.  
$H_2$. Substituing Eq.~\ref{hp} in $H_2$ yields 
\begin{eqnarray}
H_2&=&h_\bot(\sum_ib_i^+b_i+\sum_jb_j^+b_j)-B\sum_{ij}(b_i^+b_j+
b_ib_j^+)-C\sum_{ij}(b_i^+b_j^++b_ib_j),\nonumber\\
B&=&\frac{S}{2}(J_{xy}\cos{2\alpha}-J_z),\quad
C=\frac{S}{2}(J_{xy}\cos{2\alpha}+J_z).
\end{eqnarray}
The Fourier transform
$$b_{i,j}=\frac{1}{\sqrt{N}}\sum_k b_{1k,2k} \exp(i\vec k\vec r_{i,j})
$$ where $\vec k$ is a vector in reciprocal space,
leads to the momentum representation of the Hamiltonian 
\begin{eqnarray}
H_2&=&\sum_k\bigl(h_\bot(b_{1k}^+b_{1k}+b_{2k}^+b_{2k})-
2B(B_kb_{1k}^+b_{2k}+h.c.)-2C(B_kb_{1k}^+b_{2-k}^++h.c.)\bigr),
\nonumber\\
B_k&=&\exp(-i\vec k \cdot \vec c/4)\cos(\vec k \cdot \vec b/2)+ \exp(i\vec k \cdot \vec
c/4)\cos(\vec k\cdot \vec a/2).
\end{eqnarray}
The factor $B_k$  represents the $k$-dependence of the exchange interactions in
the Cu(A) subsystem. 
A separation into independent branches in the 
excitation spectrum is obtained by expressing the operators of each
sublattice with help of  collective-excitations operators $a_k $ and $c_k$ 
through 
\begin{eqnarray}
b_{1k}=\frac{1}{\sqrt{2}}(a_ke^{i\varphi_k}+c_ke^{i\psi_k})\nonumber\\
b_{2k}=\frac{1}{\sqrt{2}}(a_ke^{i\eta_k}+c_ke^{i\xi_k}),\nonumber
\end{eqnarray}
with the requirement for the phases that $\varphi_k-\eta_k=\psi_k-\xi_k=\delta$ 
and  
$\exp(2i\delta)=B_k/B_k^*$, respectively.
Using this transformation, we obtain 
\begin{eqnarray}
H_2=\sum_k\Bigl((h_\bot-2B\sqrt{B_kB_k^*})a_k^+a_k+(h_\bot+
2B\sqrt{B_kB_k^*})c_k^+c_k-\\
-2C\sqrt{B_kB_k^*}(a_ka_{-k}-c_kc_{-k}+h.c.)\Bigr).\nonumber
\end{eqnarray}
The diagonalization of the square-law part~\cite{tyablikov} gives two branches of
spin waves with Eigenenergies
\begin{equation} \label{eq:energies1}
E_k^{a,c}=\sqrt{\Bigl(h_\bot\mp
2(B+C)\sqrt{B_kB_k^*}\Bigr)\Bigl(h_\bot\mp
2(B-C)\sqrt{B_kB_k^*}\Bigr)}.
\end{equation}
{\section{Comparison with the experiment}}
For the high symmetry directions with $\vec k$ along [q,q,0], [q,q,q] and
[0,0,q] respectively,  and
in zero magnetic field (i.e. $\cos{2\alpha}=1$),  
Eq.~\ref{eq:energies1} simplifies to 
\begin{eqnarray}  
\frac{E_k^{a,c}}{h_\bot}=\sqrt{\Bigl(1\mp\frac{1}{2}\cos\frac{k_zc}{4}
(\cos\frac{k_xa}{2}+\cos\frac{k_yb}{2})\Bigr)
\Bigl(1\pm\frac{J_z}{2J_{xy}}\cos\frac{k_zc}{4}(\cos\frac{k_xa}{2}+
\cos\frac{k_yb}{2})\Bigr)}.
\label{eq:dispersion}
\end{eqnarray}
$k_{x}$, $k_{y}$ and $k_{z}$ are $\frac{2\pi}{a}h$, $\frac{2\pi}{b}k$ and 
$\frac{2\pi}{c}l$, respectively, where h,k and l are the Miller 
indices. In the isotropic case ($J_{xy}=J_{z}$), 
the two spin wave branches of Eq.~\ref{eq:dispersion} are 
degenerate. A small anisotropy lifts the degeneracy and the two 
branches split in an acoustic and an optic mode (compare 
Fig.\ref{fig:dispersion}). 
We obtain a good agreement between Eq.~\ref{fig:dispersion} and the experimental
data from a least-squares fit for the following values of the exchange parameters:
\begin{equation}
\rm nSJ_{xy}=7.7\pm0.1 \rm meV,\quad J_{xy}=45K,\quad
\frac{J_z}{J_{xy}}=0.95\pm0.01.
\label{exch}
\end{equation}
The obtained spin wave spectra are shown in Fig.~\ref{eq:dispersion}.
From our experimental data, we have only clear evidence of one 
dispersion branch (neglecting the low energy branch along c in this 
analysis). However, numerical calculations based on the Random Phase 
Approximation (RPA) method show that, at the zone center, the 
intensity of the optical branch is weak. 
Hence, due to limitations in intensity and resolution, it appears 
that exchange values given in Eq.~\ref{exch} represents an upper bound of 
anisotropy in the Cu(A) magnetic sublattice.
We note also that with these parameters the theoretical 
curves for the direction $\vec k$=(0,0,q)
lie somewhat below the experimental points (Fig.~\ref{fig:dispersion}a). This is
probably due to the fact that we neglected weak interactions betwee Cu(A) and
Cu(B), as noted above.
The interaction within the Cu(B) subsystem itself leads to a second 
phase transition toward the incommensurate magnetic phase at $T^*\sim 10K$.\par
As noted above, we observed a second excitation branch with dispersion along the (0,0,q) 
direction only. It is confined to energies much lower than the spin-wave branch 
considered here. 
This branch is probably related to spin excitations of 
the Cu(B) subsystem, which is still disordered above $T^*$. 
An analysis of the dispersion and temperature
dependences of these excitations is in progress and will be 
published later.\\
\section{Conclusion}
We have determined the spectrum of magnetic excitations in 
copper metaborate in the commensurate magnetic phase at T=12K
by means of inelastic neutron scattering. 
Two branches of excitations  have been observed with different characteristics.
Whereas the low-energy branch is attributed to fluctuations within the
Cu(B) sublattice, which only starts to order below $T^*$, the spin-wave analysis 
of the dispersion of the upper branch allowed us to determine the exchange
interactions within the Cu(A) sublattice.   
The precision of the spin model introduced here can be estimated  
by comparing the exchange parameters with values 
calculated from the molecular-field theory. This allows us
to estimate the transition temperature from the paramagnetic into the
antiferromagnetic state. In particular, if this transition is caused by the
interaction inside the Cu(A) subsystem only, the temperature of the phase
transition $T_N$ should give values for the exchange parameters which 
are close to the results of our spin-wave analysis~\cite{smart}. 
The staggered molecular field on every spin $g\mu_B(H_{e,1}S_1+H_{e,2}S_2)$ is obtained for a single
pair interaction $JS_1S_2$ from
\begin{equation}
g\mu_BH_{e,1(2)}=\frac{1}{2}\langle S_{2(1)}\rangle J.
\end{equation}
As a result, the relation between the N\'eel temperature and the exchange 
interaction 
becomes~\cite{smart} \footnote{Note, that the absence of a factor 1/2 in~\cite{smart}, as
well as in a series of others publications, leads to a decrease of the
 exchange values estimated from the N\'eel (respectively Curie) temperature.}
\begin{equation}
\frac{1}{2}\frac{JnS(S+1)}{3}=k_BT_N.
\end{equation}
For $T_N=21K $ we obtain $J=42K$. This value is close to the result
of the spin-wave analysis, which shows that the theoretical analysis of 
the spin-wave spectrum in copper metaborate in the temperature range $T^*<T<T_N$ 
can be described in a first approximation as an antiferromagnetic ordered subsystem
of Cu(A) ions with an anisotropic exchange of `easy plane' type.
Although the spins at the Cu(B) site are  still disordered above T=10K,
interactions between the two subsystems influences the excitation 
spectrum and leads to a shift of the spin-wave spectrum, in
particular along the $z$-axis.\par
This work is done under partial support from RFBR (grant
01-02-17270). We also thank E.~Kusmin, M.~Popov and A.~Pankratz for
fruitful discussions and L.P. Regnault for his help in setting 
up the experiment on IN22.

*\newpage
\end{document}